\documentclass[a4paper,11pt]{article}
\pdfoutput=1 

\usepackage{jheppub} 

\usepackage[T1]{fontenc} 
\usepackage{comment}
\usepackage{tikz}
\usepackage{amsmath,amssymb,amsfonts}
\usepackage{latexsym}
\usepackage{bbm}
\usepackage{amssymb}
\usepackage{dsfont}
\usepackage{slashed, tensor}
\usepackage{booktabs}
\usepackage{graphicx}
\usepackage{mathrsfs}

\newcommand{\be}{\begin{equation}}
	\newcommand{\ee}{\end{equation}}

\def\Q{{\mathcal{Q}}}

\newcommand{\bea}{\begin{eqnarray}}
	\newcommand{\eea}{\end{eqnarray}}
\newcommand{\ba}{\begin{array}}
	\newcommand{\ea}{\end{array}}
\newcommand{\nn}{\nonumber \\}

\title{Light-cone formalism for a point particle in a higher-spin background}

\author[a,b]{Vyacheslav Ivanovskiy}
\author[b,c]{and Dmitry Ponomarev}

\affiliation[a]{Moscow Institute of Physics and Technology,
  Dolgoprudny, 141701, Russia}
\affiliation[b]{Institute for Theoretical and Mathematical Physics,\\
Lomonosov Moscow State University,  Moscow, 119991, Russia}
\affiliation[c]{I.E. Tamm Theory Department, Lebedev Physical Institute,
 Moscow, 119991, Russia}

\emailAdd{ivanovskiy.va@phystech.edu}
\emailAdd{ponomarev@lpi.ru}

\abstract{
We study propagation of a point particle in a  massless higher-spin background employing the light-cone gauge approach. We find the  point particle action and the associated phase space Poincare charges at the leading order in higher-spin fields. We also compare our results with the analogous covariant results available in the literature.}

\begin{document} 
\maketitle
\flushbottom

\section{Introduction}

Higher-spin theories are putative theories of massless symmetric higher-spin fields. These are expected to contain a massless spin-2 field in the spectrum, which is to be identified with graviton. Massless fields inevitably require the associated  symmetries -- gauge and/or global symmetries depending on the description -- to be present in the theory. Thus, higher-spin theories are expected to provide highly symmetric extensions of gravity. It is the rich symmetry that makes higher-spin theories particularly attractive and potentially helpful in resolving various problems of modern theoretical physics with the construction of quantum gravity models among them. 

Unfortunately, constructing interacting higher-spin theories is a complicated task. As showcased by numerous no-go theorems -- see, e. g.  \cite{Weinberg:1964ew,Coleman:1967ad} 
 --  under some natural and weak assumptions massless higher-spin fields cannot interact.  
Despite these negative earlier results the quest for interacting higher-spin theories continued.  Currently, there are numerous indications that interacting higher-spin theories do exist and, moreover, few higher-spin models dealing with different simplified setups have been explicitly constructed. For a more comprehensive review on the higher-spin no-go theorems as well as on the available positive results, we refer the reader to \cite{Bekaert:2004qos,Bekaert:2010hw,Bekaert:2022poo,Ponomarev:2022vjb} and references therein.

An important feature of our Universe is that it is endowed with geometry. By this we mean that one can measure lengths, angles, volumes, study how vectors are parallel transported along trajectories, etc. All these geometric notions quantify different aspects of how point particles move in one or another background: how much proper time it takes to travel from one point to another, how trajectories bend, what tidal forces particles experience, etc. Therefore, to be able to define the geometry of one or another background, one needs to learn how point particles propagate in it.

 As is well-known, the current status on the geometry of spacetime is that it is defined exclusively in terms of a metric tensor, which at the same time serves as a gravitational field,
 the dynamical field of General Relativity. As we mentioned previously, (probably, modified) gravity is expected to form a sector in the higher-spin theory, which is, moreover, tightly intertwined with the remaining fields by higher-spin symmetries. The geometric notions, originally defined in terms of the metric are no longer invariant with respect to higher-spin transformations. This may either mean that the usual spacetime geometry cannot be made consistent with higher-spin theories or that the familiar geometric constructions need to be non-trivially extended to account for the presence of higher-spin fields. The second option is, certainly, more intriguing. A consistent higher-spin geometry may have novel interesting  features and it may, in principle, resolve  some problems of General Relativity, such as geodesic incompleteness of black hole backgrounds.

The problem of coupling of a point particle to a higher-spin background was first addressed in \cite{deWit:1979sib} by de Wit and Freedman. By considering a free higher-spin theory in the metric-like formalism and requiring the point particle action to be reparametrization invariant as well as invariant with respect to linearized higher-spin gauge symmetries, they found the point particle action at the leading order in higher-spin fields. An analogous result can be formulated using the frame-like language, see \cite{Tarusov:2023rad}.

Then, in  \cite{Segal:2000ke,Segal:2001di,Segal:2002gd}\footnote{See also \cite{Bars:2001um} for a closely related result.} Segal suggested  that the general reparametrization invariant point particle action in the Hamiltonian form should be interpreted as the action of a point particle propagating in a conformal higher-spin background. 
This suggestion is based on a simple observation that, once the point-particle Hamiltonian is properly identified with higher-spin fields, linearised symmetries of general reparametrization invariant point particle dynamics   coincide with free conformal higher-spin  symmetries. This line of thought eventually resulted in a fully non-linear conformal higher-spin theory \cite{Segal:2002gd}, which was also constructed using alternative considerations in \cite{Tseytlin:2002gz}.
However, it needs to be remarked that the conformal higher-spin theory possesses a symmetry, which is a deformed version of the symmetry of the original point-particle theory.
Accordingly, the point-particle dynamics of \cite{Segal:2000ke} is consistent in non-linear conformal higher-spin backgrounds only up to certain corrections\footnote{More precisely, the original symmetry results in a conformal higher-spin theory with the action which is free of space-time derivatives. To remedy this problem, it was suggested to deform the conformal higher-spin theory by 
replacing the usual products of fields with the Moyal products. This deformation does make the conformal higher-spin dynamics non-trivial, however, it also deforms higher-spin symmetries with higher-derivative corrections.}.

Higher-spin geometry was also discussed in the context of other higher-spin theories, in particular, for the Chern-Simons theories \cite{Blencowe:1988gj} and for the Vasiliev theory \cite{Vasiliev:1990en,Vasiliev:2003ev}. In these theories, the traditional geometric notions based on metric, indeed, turn out to be not higher-spin invariant \cite{Didenko:2009td,Ammon:2011nk,Iazeolla:2011cb,Castro:2011fm,Ammon:2012wc,Ammon:2013hba,Didenko:2021vui,Iazeolla:2022dal}. Despite some general suggestions on how this problem can be addressed were made\footnote{In particular,   propagation of scalar fields in certain  backgrounds in Chern-Simons higher-spin theories was studied in \cite{Kraus:2012uf} . This approach is analogous to ours, except that we use point particles instead of scalar fields as probes of geometry.} higher-spin geometry in these theories remains poorly understood.

In the present paper, we initiate the analysis of point particles propagating in chiral higher-spin backgrounds. 
 A major advantage of chiral higher-spin theories \cite{Metsaev:1991mt,Metsaev:1991nb,Ponomarev:2016lrm}\footnote{Chiral higher-spin theories were suggested in \cite{Ponomarev:2016lrm}. This suggestion heavily relies on the earlier analysis by Metsaev \cite{Metsaev:1991mt,Metsaev:1991nb}. Another closely related result can be found in \cite{Devchand:1996gv}. For recent developments on chiral higher-spin theories, see \cite{Ponomarev:2017nrr,Skvortsov:2018jea,Skvortsov:2020wtf,Skvortsov:2020gpn,Krasnov:2021nsq,Sharapov:2022faa,Ponomarev:2022atv,Monteiro:2022lwm,Bu:2022iak,Tran:2022tft,Ponomarev:2022ryp,Ponomarev:2022qkx,Herfray:2022prf,Adamo:2022lah,Monteiro:2022xwq}.} is that these are formulated in a closed and very compact form.
In particular, the associated action truncates at the cubic order in fields. Considering the simplicity of the chiral higher-spin theory, one may hope that the point particle  action in chiral higher-spin backgrounds can either be found in a closed form as well or, at least, it can be systematically studied at the leading orders in fields. 
In the latter case, one may expect that already this leading order analysis will be able to answer important qualitative questions, such as whether a suitable point-particle action  exists and, if does, how many independent coupling constants it features. 

Another significant advantage of chiral higher-spin theories in the present context is that the associated equations of motion can be solved exactly in a systematic manner. This property is related to the fact that chiral higher-spin theories are natural higher-spin generalisations of self-dual Yang-Mills and self-dual gravity \cite{Ponomarev:2017nrr,Krasnov:2021nsq} and similarly to the latter theories, these are integrable. Integrability, in turn, is related to the existence of abundant exact solutions  
\cite{mason1996integrability,hitchin1999integrable}. By applying the known solution-generating techniques to the chiral higher-spin case one should be able to find exact  chiral higher-spin backgrounds and then study their geometric properties. 

It is worth emphasising that similarly to self-dual Yang-Mills and self-dual gravity, chiral higher-spin theories have the action which is not real and, thus, these cannot be regarded as a fully satisfactory solution to the original higher-spin interaction problem. It would be, certainly, very interesting to find a parity-invariant completion of chiral higher-spin theories. At the same time,  
 while such a completion is not yet available (it is not  known whether it exists; recent developments in this direction can be found in \cite{Ponomarev:2022atv,Tran:2022tft,Ponomarev:2022ryp,Ponomarev:2022qkx,Adamo:2022lah}), we may still explore geometric aspects of a simpler chiral theory expecting that these already capture characteristic  features  common to  higher-spin theories in general.
 
 More specifically, in the present paper we construct the action of the point particle in the higher-spin background at the leading order in higher-spin fields. We do that by applying the standard light-cone formalism to the point-particle case. At the given order in perturbations interactions of the chiral higher-spin theory do not yet contribute.
 Our result is, therefore, analogous to the covariant one of  \cite{deWit:1979sib}, except that we work in the light-cone formalism, which is more adapted to dealing with the complete chiral higher-spin theories in future. In the course of our analysis, we also clarify various subtleties related to the application of the light-cone approach to the point-particle case. In particular, we extend the standard discussion of fake interactions to point particles as well as explain how particle's back-reaction to the background can be consistently removed.
 We also compare our action with the leading order covariant results  \cite{deWit:1979sib,Segal:2000ke}.

The remaining part of the paper is organized as follows. We start by recalling the general ideas behind the light-cone formalism and then review the relevant description of free massless higher-spin fields in 4d Minkowski space in section \ref{s2}.
Next, in section \ref{secfpp} we extend this approach to a free point particle. Then, in section \ref{s3} we discuss how the light-cone formalism can be used to construct a theory of a point particle interacting with a higher-spin background. This general discussion is followed by section \ref{s4}, in which we construct the point particle action at the leading order in fields. We then conclude in section \ref{sec:5}.  In appendices we give our conventions as well as compare our result with its covariant counterpart.

\section{The light-cone formalism and free massless fields \label{s2}}

The key feature of the light-cone gauge approach to massless fields is that it only deals with the physical degrees of freedom. This, however, implies that unlike in covariant approaches Lorentz invariance is no longer manifest. Instead, it should be imposed manually. In practice, this is, usually, done following \cite{Dirac:1949cp}. More specifically, any Poincare invariant theory by the Noether theorem has a set of conserved charges, with one charge associated with each generator of the Poincare algebra. In the Hamiltonian formalism these charges generate the action of the Poincare algebra on theory's phase space via the Dirac bracket. 
This entails that the conserved charges commute with the Dirac bracket the same way as the generators of the Poincare algebra associated with these charges do\footnote{The approach of \cite{Dirac:1949cp} is, in fact,  a bit more general. Namely, it does not rely on the existence of an action as well as it does not rely on any particular choice of a time direction. As a result, the analysis features few Hamiltonians -- also known as dynamical generators -- that enter on equal footing. The only requirement imposed in \cite{Dirac:1949cp} is that there is a set of properly commuting phase space charges.}. 

This general idea can be used to construct interacting theories possessing Poincare symmetry in the following way. One starts with a Poincare invariant free theory and extracts the associated conserved charges. For interacting theories these charges receive non-linear corrections. By requiring that the deformed charges of the non-linear theory commute as the associated generators of the Poincare algebra, one obtains a series of constraints, which can be solved order by order in perturbations.

The approach  outlined above was used to construct cubic vertices for massless higher-spin fields \cite{Bengtsson:1983pd,Bengtsson:1986kh} and, eventually, resulted in the chiral higher-spin theory \cite{Metsaev:1991mt,Metsaev:1991nb,Ponomarev:2016lrm}. For a review on the light-cone formalism in this context we refer the reader to \cite{Ponomarev:2022vjb}. Below we will use this formalism to construct the leading order correction to the action of a point particle propagating in the higher-spin background. In the present section, we will start by briefly reviewing the light-cone formalism for free massless fields in  Minkowski space. We will then present an analogous construction for a free particle in  Minkowski space in the next section.

	\subsection{Free massless fields in the light-cone gauge}

	The Poincare algebra commutation relations are given by
	\begin{equation}
		\begin{split}
		\label{1jun1}
		[P^a,P^b] &\,= 0,\\
		[J^{ab},P^c] &\,= P^a \eta^{bc}- P^b \eta^{ac},\\
		[J^{ab}, J^{cd}] &\, = J^{ad} \eta^{bc} -J^{bd} \eta^{ac} -J^{ac} \eta^{bd}+ J^{bc} \eta^{ad},
		\end{split}
	\end{equation}
	where $x^a=\{x^-,x^+,x,\bar x\}$. For details on our conventions we refer the reader to appendix  \ref{App}. 
	
	Commutation relations (\ref{1jun1}) admit helicity-$\lambda$ representations
\begin{equation}
		\begin{split}
		P^a \cdot \Phi^{\lambda}&\,\equiv \partial^a \Phi^{\lambda},\\
		\label{1.1}
		J^{ab}\cdot \Phi^{\lambda}&\,\equiv (x^a\partial^b - x^b \partial^a+S^{ab} )\Phi^\lambda,
		\end{split}
	\end{equation}
where $S^{ab}$ is the spin part of the angular momentum and it is  given by
\begin{equation}
\begin{split}
\label{1.3}
S^{+a}\cdot \Phi^{\lambda} &=0,\qquad S^{x\bar x} \cdot \Phi^{\lambda} = -\lambda \Phi^{\lambda},\\
S^{x-}\cdot \Phi^{\lambda} &= \lambda  \frac{\partial}{\partial^+}\Phi^{\lambda}, \qquad 
		S^{\bar x -}\cdot \Phi^{\lambda}= -\lambda   \frac{\bar\partial}{\partial^+}\Phi^{\lambda}.
\end{split}
\end{equation}

	In light-cone gauge the free action for a set of massless fields of helicities 
	$\lambda$ is
	\begin{equation}
		\label{1.6}
		S_2\equiv \int d^4 x L_2, \qquad L_2 = -\frac{1}{2}\sum_{\lambda}  \partial_a\Phi^{-\lambda} \partial^a \Phi^\lambda.
	\end{equation}
	Here we do not make any assumptions on the spectrum of values $\lambda$ takes except that opposite helicities enter in pairs.

In the light-cone formalism it is convenient to take $x^{+}$ as a time variable. Accordingly, the canonical momentum is given by
	\begin{equation}
	\label{1jun2}
		\Pi^\lambda \equiv \frac{\delta L_2}{\delta( \partial^- \Phi^\lambda)}= -\partial^+ \Phi^{-\lambda}.
	\end{equation}
	Then, the Hamiltonian reads
	\begin{equation}
		\label{1.8}
		H^\Phi_2\equiv  \sum_{\lambda }  \int  d^{3}x^\perp(\Pi^\lambda \partial^-\Phi^{\lambda} - L_2)  =
		\sum_{\lambda} \int d^{3}x^\perp \partial \Phi^{-\lambda} \bar\partial \Phi^\lambda,
	\end{equation}
	where $x^{\perp}\equiv \{x,\bar x,x^- \}$  and integration goes over equal-time hypersurfaces.

Due to the fact that the Lagrangian (\ref{1.6}) is of first order in time derivatives (\ref{1jun2}) presents a constraint.
As a consequence, the usual Poisson bracket  should be replaced with the appropriate  Dirac bracket. The  analysis of constraints and the derivation of the appropriate Dirac bracket in this case is very standard -- see, e.g. \cite{Heinzl:2000ht} --, so we just give the end result
\begin{equation}
		\label{1.9}
		[\partial^+ \Phi^\lambda(x^\perp,x^+),\Phi^{\mu}(y^\perp,x^+)]_\Phi= \frac{1}{2}\delta^{\lambda+\mu,0}\delta^{3}(x^\perp,y^\perp).
	\end{equation}
Equivalently, (\ref{1.9}) can be rewritten as
	\begin{equation}
		\label{1.10}
		[\Phi^\lambda(x^\perp,x^+),\Phi^{\mu}(y^\perp,x^+)]_\Phi= \frac{1}{\partial^+_x - \partial^+_y}\delta^{\lambda+\mu,0}\delta^{3}(x^\perp,y^\perp),
	\end{equation}
	where the subscript ''$\Phi$'' means that $[\cdot,\cdot]_\Phi$ acts on fields, as opposed to phase space variables of a point particle, which will appear below.

	The canonical Hamiltonian (\ref{1.8}) and the Dirac bracket (\ref{1.10}) define the time evolution in the standard way
	\begin{equation}
		\label{1.11}
		\partial^- F(\Phi) = [F(\Phi), H^\Phi_2]_\Phi.
	\end{equation}
It is easy to see that (\ref{1.11}) agrees with the equation of motion resulting from (\ref{1.6}). This serves as a simple consistency test of the Dirac bracket (\ref{1.10}).

\subsection{Noether currents and charges.} 
The action (\ref{1.6}) exhibits invariance under the Poincare algebra transformations (\ref{1.1}). This entails conservation of the associated Noether currents
	\begin{align}
		\notag
		P^i \quad \to  \;\quad T^{i,j} &\,= \sum_{\lambda}\frac{\delta L_2}{\delta (\partial_j \Phi^\lambda)}\partial^i \Phi^\lambda - \eta^{ij} L_2,\\
		J^{ij} \quad   \to  \quad L^{ij,k} &\,= x^i T^{j,k}-x^j T^{i,k}+R^{ij,k},
		\label{1.12}
	\end{align}
	where $R^{ij,k}$ is the spin current
	\begin{equation}
		\label{1.13}
		R^{ij,k} \equiv \sum_{\lambda}\frac{\delta L_2}{\delta( \partial_k \Phi^\lambda)} S^{ij}\cdot \Phi^\lambda
	\end{equation}
	and  $S^{ij}$ was given in (\ref{1.3}).

 The Noether charges are defined in the standard way
	\begin{equation}
		\label{1.14}
		Q_2[P^i] \equiv \int d^3x^\perp T^{i,+}, \qquad Q_2[J^{ij}] \equiv \int d^3x^\perp L^{ij,+}.
	\end{equation}
	Here  the subscript "2" refers to the fact  that these charges are quadratic in fields. It is convenient to choose the integration hypersurface at $x^+=0$.
Evaluating  the charges (\ref{1.14}) explicitly, one finds 
	\begin{equation}
		\label{1.15}
		Q_2[P^i] = -\sum_{\lambda}\int d^3x^\perp \partial^+\Phi^{-\lambda}  p_2^i \Phi^\lambda, \qquad
		  Q_2[J^{ij}] = -\sum_{\lambda}\int d^3x^\perp \partial^+\Phi^{-\lambda}  j_2^{ij} \Phi^\lambda,
	\end{equation}
	where 
	\begin{align}
		\notag
		p_2^+ &\,=\partial^+, & p_2^- &= -\frac{\partial \bar\partial}{\partial^+}, \qquad\qquad \quad\;  p_2 = \partial,  \qquad\qquad\quad\;    \bar p_2=
		\bar\partial,\\
		\notag
		j_2^{+-} &\,=  - x^-\partial^+,&  j_2^{x\bar x} &= x\bar\partial - \bar x \partial -\lambda, \\
		\notag
		j_2^{x+}&\, = x\partial^+, &  j_2^{x-} &= -x\frac{\partial \bar\partial}{\partial^+} - x^- \partial+\lambda\frac{\partial}{\partial^+}, \\
		j_2^{\bar x+}&\, = \bar x\partial^+,
		&  j_2^{\bar x-} &= -\bar x\frac{\partial \bar\partial}{\partial^+} - x^- \bar\partial
		-\lambda\frac{\bar\partial}{\partial^+}.
		\label{1.16}
	\end{align}

	As expected,  the charges (\ref{1.15}) generate the action of the Poincare algebra 
	\begin{equation}
		\label{1.17x1}
		[\Phi^\lambda, Q_2[P^i]]_\Phi = p_2^i \Phi^\lambda, \qquad  [\Phi^\lambda, Q_2[J^{ij}]]_\Phi = j_2^{ij} \Phi^\lambda.
	\end{equation} 
	This action is realized on phase space associated with the Cauchy surface $x^+=0$.
	Once time derivatives are eliminated from the original representation (\ref{1.1}) via equations of motion, it coincides with (\ref{1.17x1}) on $x^+=0$, as required.
	Moreover, the charge $P^-_2$ associated with the light-cone time translation is just the canonical Hamiltonian
	$H^\Phi_2$  (\ref{1.8}).

	\section{Free point particle}
	\label{secfpp}
	
	The action for a free point particle in Minkowski space is
	\begin{equation}
	\label{2jun1}
		S=-m\int d \tau \sqrt{-\eta_{\mu\nu}\dot x^{\mu} \dot x^{\nu}}=-m\int d \tau \sqrt{-2\dot x^{+}\dot x^- -2\dot x \dot{\bar x}},
	\end{equation}
 where $\dot x$ denotes the derivative with respect to the world-line parameter $\tau$, and $m$ is particle's mass. This action exhibits reparametrization invariance. 
 The key feature of the light-cone formalism is that one fixes all gauge ambiguities at the expense of breaking manifest Lorentz invariance. For the point particle reparametrization invariance is an ambiguity of this type.
  It seems natural to fix it coherently with the analysis in the field theory case.
  More specifically, we set the world-line parameter equal to the light-cone time,
  $\tau=x^+$. This leads us to the action
	\begin{equation}
		S=-m\int dx^{+}\sqrt{-2\dot x^- -2\dot x \dot{\bar x}}.
		\label{action}
	\end{equation}
As in the field theory case, $x^{+}$ is regarded as the  time coordinate, while the other three coordinates $\{x,\bar x,x^- \}$ are treated as space coordinates.

	The canonical momenta for (\ref{action}) are defined as usual
\begin{equation}
	\begin{split}
		p_-\equiv \frac{\partial L}{\partial \dot x^-}=\frac{m}{ \sqrt{-2\dot x^- -2 \dot x { \dot{\bar{x}}}}},\\
		p_x\equiv \frac{\partial L}{\partial \dot x}=\frac{m { \dot{\bar{x}}}}{ \sqrt{-2\dot x^- -2 \dot x { \dot{\bar{x}}}}},\\
		p_{\bar x}\equiv \frac{\partial L}{\partial \dot x}=\frac{m { \dot{{x}}}}{ \sqrt{-2\dot x^- -2 \dot x { \dot{\bar{x}}}}}.
	\end{split}
\label{momenta}
\end{equation}
Solving (\ref{momenta}) for velocities, we find
\begin{equation}
\label{7jun11}
	{ \dot{\bar{x}}}=\frac{p_x}{p_-}, \qquad \dot x=\frac{p_{\bar x}}{p_-},\qquad \dot x^- = -\frac{p_x p_{\bar x}}{p_-^2}-\frac{m^2}{2p_-^2}.
\end{equation}
  This allows us to find the point-particle Hamiltonian 
  \begin{equation}
  	\label{Ham}
  	H_p\equiv p_-\dot x^-+p_{x}\dot x + p_{\bar x}\dot{\bar{x}}-L = \frac{p_x p_{\bar x}}{p_-}+\frac{m^2}{2p_-}.
  \end{equation}
  Finally,  the Poisson bracket has the standard form
\begin{equation}
	[f, g]_p=\frac{\partial f}{\partial x^i}\frac{\partial g}{\partial p_i}-\frac{\partial f}{\partial p_i}\frac{\partial g}{\partial x^i},
\end{equation}
where the subscript ''$p$'' refers to the fact that the bracket  $[\cdot,\cdot]_p$ acts on coordinates and momenta of the point particle.
	
	 \subsection{Noether currents and charges} 
The action (\ref{action}) is invariant with respect to transformations from the Poincare algebra, which act as follows
\begin{equation}
J^{ab}[x^c]=x^a\eta^{bc}-x^b \eta^{ac}, \qquad P^a[x^b]=\eta^{ab}.
\label{inv}
\end{equation}  
According to the Noether theorem, each infinitesimal symmetry transformation  
\begin{equation}
	\label{1.17}
	\delta x^i = \delta\omega \frac{\delta x^i}{\delta \omega}, \qquad \delta t = \delta\omega \frac{\delta t}{\delta \omega},
\end{equation}
where $\delta\omega$ is an infinitesimal transformation parameter, entails the presence of a conserved quantity
 \begin{equation}
 	\label{1.26}
 	j = H \frac{\delta t}{\delta \omega}-p_{i}\frac{\delta x^i}{\delta \omega}.
 \end{equation}
By reading off $\delta x/\delta \omega$ and $\delta t/\delta\omega$ for each infinitesimal Poincare algebra transformation from (\ref{inv}) and applying (\ref{1.26}), we find the following list of conserved quantities for (\ref{action})
\begin{equation}
	\label{1.28}
	\begin{split}
		q_0 [P^x] = -p^x, \qquad q_0 [P^{\bar x}] = -p^{\bar x}, \qquad q_0 [P^+] = -p^+, \qquad q_0 [P^-] = H_p,\\
		q_0 [J^{x\bar x}] = \bar xp^x-x p^{\bar x}, \qquad q_0 [J^{x+}] = -xp^+, \qquad 
		q_0 [J^{\bar x+}] = -\bar xp^+,\\
		q_0 [J^{x-}] = H_p x + p^x x^-, \qquad  q_0 [J^{\bar x-}] = H_p \bar x + p^{\bar x} x^-,
		\qquad  q_0 [J^{+-}] =  p^+ x^-.
	\end{split}
\end{equation}
	Here, for simplicity, we  deal with the point-particle phase space on the Cauchy surface at $x^+=0$, so all terms proportional to $x^+$ were dropped.
	By construction, 
	conserved quantities (\ref{1.28}) generate the action of the Poincare algebra on the aforementioned phase space. As a consistency check, we verified that for any pair of the Poincare algebra generators $O^{(1)}$ and $O^{(2)}$ one has
	\begin{equation}
	[q_0[O^{(1)}], q_0[O^{(2)}]]_p=q_0[[O^{(1)},O^{(2)}]].
\end{equation}

\section{Generalities on interactions \label{s3}}

In the previous sections we reviewed how the free massless higher-spin fields as well as the free point particle in  Minkowski space are described in the light-cone formalism. In the present section we extend this discussion to the case of the point particle propagating in the higher-spin background. More specifically, below we require that the aforementioned system possesses Poincare invariance and derive the consistency conditions it entails.
  In the next section we will solve the associated constraints at the leading order in higher-spin fields.

\subsection{Phase space, the bracket and the charges} 
Phase space of a joint system consisting of a point particle and of higher-spin fields is just the tensor product of phase spaces of the two separate theories.  Coordinates on this space are 
\begin{equation}
\label{6jun1}
(\Phi(z^{\perp}),x^{\perp}, p_{\perp}),
\end{equation}
where from now on we use $z^\perp$ for transverse coordinates on the field theory side to distinguish them from transverse coordinates of a point particle, $x^\perp$. The Dirac bracket for the joint system we are dealing with is the sum of the Dirac brackets for fields and for a point particle separately
\begin{equation}
	[\cdot, \cdot] \equiv [\cdot, \cdot]_p+[\cdot, \cdot]_\Phi.
\end{equation} 

The Poincare charges in the interacting theory are functions/functionals on the joint phase space. The natural physical requirement is that  charges responsible for interactions of a particle with higher-spin fields are local in the sense that a particle only experiences values of fields and their derivatives at the point it is located. For example, a local charge, which is linear in fields has the following form
\begin{equation}
\label{6jun2}
k(x^{\perp}, p_{\perp},\partial_\perp)\Phi^{\lambda}(x^{\perp}).
\end{equation}

For a charge  (\ref{6jun2}) it is clear that $[\cdot, \cdot]_\Phi$ only acts on the field. At the same time, it may be somewhat confusing whether $[\cdot, \cdot]_p$ should act on the $x^\perp$ argument of $\Phi$ or not. To answer this question, we rewrite (\ref{6jun2}) as
\begin{equation}
	k(x^{\perp}, p_{\perp},\partial_\perp)\Phi^{\lambda}(x^{\perp})=\int d^{3}z^\perp \delta(x^\perp-z^{\perp}) k(x^{\perp}, p_{\perp},\partial_{z^\perp})\Phi^{\lambda}(z^{\perp}) .
	\label{2.2}
\end{equation}
In this form we manifestly isolated our field phase space variable $\Phi^\lambda(z^\perp)$. As explained above, the point-particle Poisson bracket $[\cdot, \cdot]_p$ only acts on particle's phase space variables. Accordingly, it does not act on $\Phi^\lambda(z^\perp)$. At the same time, it acts both on $k$ and the delta function, as $k$ depends on $x^\perp$ and $p_\perp$, while the delta function features $x^\perp$ in the argument.
To understand how this translates to the action on charges in the form (\ref{6jun2}) we just need to integrate out $z^\perp$. As it is not hard to see, 
the fact that $[\cdot, \cdot]_p$ acts on the delta function in (\ref{2.2}) implies that it also acts on the $x^\perp$ argument of $\Phi^\lambda$ in (\ref{6jun2}).

The presence of the second term -- with the argument of the field transformed -- can be explained intuitively as follows. Transformations in the point particle phase space, certainly, have no effect on the phase space of fields. However, the charge (\ref{6jun2}) has a particular form as it depends only on the values of the field at the location of the point particle. As a result, once the location of the point particle changes, the field needs to be evaluated at a different point. This, in effect, means that for charges like (\ref{6jun2}) the point particle phase space transformations also act on the $x^\perp$ argument of the field as well.

\subsection{Interactions in the light cone approach}

We have already outlined above how the light-cone formalism allows one to construct consistent interacting theories in general. We will now focus on the application of this procedure to the case of a point particle in the field theory background.

As usual, in the interacting theory the Poincare charges of the free theory get deformed with non-linear terms. In our case the charge associated with the Poincare algebra generator $O$ in the interacting theory admits the following expansion
\begin{equation}
	\label{2.3}
	\begin{split}
	\mathcal Q [O]& \equiv Q[O]+q[O], \\
	Q [O] &\equiv  Q_2[O]+Q_3[O] +\dots, 
\\
 q[O] &\equiv 
q_0[O]+q_1[O]+q_2[O]+\dots.
\end{split}
\end{equation}
Here $Q_2$ and $q_0$ are the charges of the free field theory and of the free point particle that we found in the previous sections. The remaining terms represent non-linear corrections. In particular, $Q_3$ is the field theory charge, which is cubic in  fields. In (\ref{2.3}) it is followed by $\dots$, which refer to  higher-order field theory charges, such as $Q_4$, $Q_5$ etc. These may, in principle, be present, however, they are absent in the chiral higher-spin theory case. The remaining charges in (\ref{2.3}) correspond to interactions of the point particle with the field. In particular, $q_1$ is linear in fields, $q_2$ is quadratic, etc.

The key consistency requirement that interactions need to satisfy is that they do not violate the Poincare invariance of the free theory. The standard argument then allows one to conclude that the Poincare invariance of the interacting theory implies that for any pair of the Poincare algebra  generators $O^{(1)}$ and $O^{(2)}$, such that
\begin{equation}
\label{6jun3}
[O^{(1)},O^{(2)}]=O^{(3)},
\end{equation}
the associated Poincare charges should commute accordingly
\begin{equation}
	\label{2.5}
	[\Q^{(1)},\Q^{(2)}]=\Q^{(3)}.
\end{equation}
Here we used the shorthand notation $\Q[O^{(i)}]\equiv \Q^{(i)}$. By splitting the charges as well as the Poisson bracket into contributions associated with the fields and with the particle, we find
\begin{equation}
\label{6jun4}
	[Q^{(1)},Q^{(2)}]_\Phi+[q^{(1)},q^{(2)}]_p + [q^{(1)},q^{(2)}]_\Phi +[q^{(1)},Q^{(2)}]_\Phi + [Q^{(1)},q^{(2)}]_\Phi =Q^{(3)}+ q^{(3)}.
\end{equation} 
Considering that the field theory alone is Poincare invariant, that is
\begin{equation}
[Q^{(1)}, Q^{(2)}]_\Phi=Q^{(3)},
\end{equation}
we can simplify (\ref{6jun4}) to
\begin{equation}
	[q^{(1)},q^{(2)}]_p + [q^{(1)},q^{(2)}]_\Phi +[q^{(1)},Q^{(2)}]_\Phi + [Q^{(1)},q^{(2)}]_\Phi = q^{(3)}.
	\label{3.2}
\end{equation}
This gives the consistency condition for the point particle interacting with the background fields.

\subsection{Removing back-reaction} 

An important part of the setup in which a point particle propagates in some background is that the background is unaffected by the presence of the particle. In the previous discussion, instead, we rather considered a system of a point particle interacting with a field, that is not only the field affects the motion of the particle, but the particle also sources fields back. Thus, to describe the desired system, we need to correct the above analysis, by removing terms, which are responsible for particle's back-reaction. 

 It is also worth remarking that the presence of the back-reaction leads to well-known technical difficulties. Namely, the field  sourced by the point particle is infinite at the location of the particle.
 As a result, say, the contribution to the particle's action from its interaction with the field it sources is infinite. The divergence of electron's self-energy in classical field theory is a famous example of this phenomenon.
 It is not hard to see that infinite terms are, indeed, present in our analysis. Namely, by using arbitrary local charges $q^{(1)}$ and $q^{(2)}$ and evaluating their bracket $[q^{(1)},q^{(2)}]_\Phi$, one inevitably encounters divergence  $\delta(0)$.

 Below we will show how the back-reaction can be consistently removed and find the associated version of the consistency condition  (\ref{3.2}). The latter will be  free of any divergent terms.

To understand how back-reaction can be consistently removed, we need to go  one step back and recall how the light-cone consistency conditions, such as (\ref{2.5}), are derived. One starts by considering the action of a given charge on any function $f$ of phase space variables 
\begin{equation}
\label{6jun6}
	\delta f = [f, \Q[O]].
\end{equation}
For a set of generators that satisfy (\ref{6jun3}) consistency requires that the associated transformations of $f$ satisfy 
\begin{equation}
\label{6jun5}
[\delta^{(1)}, \delta^{(2)}]f=\delta^{(3)}f.
\end{equation}
 Employing (\ref{6jun6}) and the Jacobi identity for the left-hand side of (\ref{6jun5}) we find
\begin{equation}
	[\delta^{(1)}, \delta^{(2)}]f=\left[\left[f,\Q^{(2)}\right], \Q^{(1)}\right]-\left[\left[f, 
	\Q^{(1)}\right], \Q^{(2)}]\right]=\left[f,\left[\Q^{(1)}, \Q^{(2)}\right]\right].
	\label{3.7}
\end{equation}
At the same time, the right-hand side (\ref{6jun5}) gives $[f, \Q^{(3)}]$. Considering that it should be equal (\ref{3.7}) for any $f$, we derive (\ref{2.5}).

To understand how the above analysis should be changed once the back-reaction is removed, we decompose (\ref{6jun6}) as
\begin{equation}
	\delta f = [f,Q+q]= [f,q]_p +[f,q]_\Phi + [f,Q]_\Phi.
	\label{3.3}
\end{equation}
The term $[f,Q]_p$ is absent on the right-hand side of (\ref{3.3}) because $Q$ does not depend on the coordinates and momenta of the particle.
 Phase space coordinates themselves can be regarded as particular phase space functions $f$. From (\ref{3.3}) these transform as
\begin{equation}
	\delta \Phi =  [\Phi,Q]_\Phi +[\Phi,q]_\Phi, \qquad \delta x^\perp = [x^\perp,q]_p, \qquad \delta p^\perp = [p^\perp,q]_p.
	\label{3.4}
\end{equation}

 It is not hard to see that the second term in the transformation for $\Phi$ in (\ref{3.4}) is the back-reaction term, as it involves the point-particle charge $q$.
Besides that, it is supported at the location of the point particle, which is how the back reaction is supposed to act. The remaining terms, in turn, represent transformations of fields in the field theory alone, as well as the action of the field background on the point particle. These latter terms are not related to back-reaction and, thus,  should be kept. 
Returning to the general formula (\ref{3.3}), we conclude that the second term on the right-hand side is the back-reaction term.

We have identified so far that removing particle's back-reaction amounts to removing $[\Phi,q]_\Phi$ in (\ref{3.4}). Our goal now is to find out how this affects the consistency condition on the commutators of charges (\ref{3.2}). This can be done using the following shortcut. One starts by considering a rescaling 
\begin{equation}
	x^\perp \to \epsilon x^\perp, \qquad p_\perp \to \epsilon p_\perp, \qquad [,]_p \to \epsilon^{-1}[,]_p,
\end{equation}
which converts (\ref{3.4}) into 
\begin{equation}
	\delta \Phi^\lambda =  [\Phi^\lambda,Q]_\Phi +\epsilon[\Phi^\lambda,q]_\Phi, \qquad \delta x^\perp = [x^\perp,q]_p, \qquad \delta p^\perp = [p^\perp,q]_p.
	\label{3.9}
\end{equation}
By applying the same rescaling to (\ref{3.2}) one finds 
\begin{equation}
	\epsilon [q^{(1)},q^{(1)}]_p +\epsilon^2 [q^{(1)},q^{(2)}]_\Phi +\epsilon [q^{(1)},Q^{(2)}]_\Phi + \epsilon [Q^{(1)},q^{(2)}]_\Phi = \epsilon q^{(3)}.
	\label{3.10}
\end{equation}
To remove the back-reaction term in (\ref{3.9}) it suffices to send $\epsilon$ to zero. In this limit (\ref{3.10}) leads to 
\begin{equation}
	[q^{(1)},q^{(2)}]_p  +[q^{(1)},Q^{(2)}]_\Phi + [Q^{(1)},q^{(2)}]_\Phi = q^{(3)},
	\label{3.8}
\end{equation}
which represents the consistency condition for a system  with back-reaction removed.

One may be rightfully worried that (\ref{3.8}) was obtained by employing a potentially singular limit $\epsilon\to 0$ and, thus, may not be valid. To resolve these concerns we checked explicitly that 
(\ref{3.9}) with $\epsilon=0$ does lead to (\ref{3.8}). To show this, one needs to follow the steps that lead  from (\ref{6jun6}), (\ref{3.3}) to (\ref{2.5}) detailed above, but with the back-reaction term -- the second term on the right-hand side of (\ref{3.3}) -- removed.

\subsection{Kinematical and dynamical constraints} 

In the light-cone approach there are naturally two types of generators. The generators of the first type leave the Cauchy surface -- $x^+=0$ in our case -- invariant. These generators do not involve time evolution and, as a result, remain undeformed in the interacting theory \cite{Dirac:1949cp}. Generators of this type are called kinematical\footnote{There are exotic theories, for which kinematical generators get deformed \cite{0521670535}.}. The second type of generators, instead, is transverse to the Cauchy surface, thus, they involve time evolution. These get deformed in the interacting theory and are called dynamical.
For our choice of the Cauchy surface only $P^-$, $J^{x-}$ and $J^{\bar x -}$ are dynamical generators, while the remaining ones are kinematical. 
It is conventional to collectively denote kinematical and dynamical generators  as $K$ and $D$ respectively. 

The splitting of the Poincare algebra generators into kinematical and dynamical ones leads to the splitting of the Poincare algebra commutators into three groups depending on the types of generators these commutators involve. The first group is, schematically, of the form
\begin{equation}
	\label{2.8}
	[K^{(1)},K^{(2)}]=K^{(3)}.
\end{equation}
Considering that kinematical generators remain undeformed in the non-linear theory, the constraint on charges associated with (\ref{2.8}) is automatically satisfied in the non-linear theory as a consequence of it being true in the free one. 

The second group of commutators is of the form 
\begin{equation}
	[K^{(1)},D] = K^{(2)}, \qquad [K,D^{(1)}]=D^{(2)}.
\end{equation}
The associated constraints for charges are referred to as the kinematical constraints. The key feature of these constraints is that these are linear in unknown deformations of $\Q[D]$. As a result, these can be solved systematically at all orders in fields. 

Finally, the last group of commutators is of the form 
\begin{equation}
	[D^{(1)}, D^{(2)}]=0.
	\label{3.14}
\end{equation}
The associated constraints are referred to as dynamical constraints. These involve deformations of $\Q[D]$ quadratically and, due to that they are quite challenging to solve. 

As a final remark, we remind that the time translation generator is the Hamiltonian
\begin{equation}
\label{6jun8}
H_p=q[P^-].
\end{equation}
Thus, once a set of consistent charges in the interacting theory is found the point-particle action can be obtained from the standard formula
\begin{equation}
\label{6jun9}
S= \int(  p_-\dot x^-+p_{x}\dot x + p_{\bar x}\dot{\bar{x}} - H_p)dx^+.
\end{equation}

\section{Solution at the leading order \label{s4}}

With all the preliminary work done, we are ready to proceed to the solution for the leading order charges $q_1$. To this end, we consider (\ref{3.8}) and keep only the leading order contribution in fields. This gives 
\begin{equation}
	\label{2.6}
	\begin{split}
	[Q_2[O^{(1)}],q_1[O^{(2)}]]_\Phi+[q_1[O^{(1)}],Q_2[O^{(2)}]]_\Phi+
	[q_0[O^{(1)}],q_1[O^{(2)}]]_p\\
	 +[q_1[O^{(1)}],q_0[O^{(2)}]]_p
	=q_1[O^{(3)}].
	\end{split}
\end{equation}

The most general deformation of the dynamical generators at the given order reads 
\begin{eqnarray}
	&&q_1[P^-]=\sum\limits_\lambda h^{\lambda}(x^{\perp}, p_{\perp}, \partial_{x^{\perp}}) \Phi^{\lambda}(x^{\perp}),
	\label{2.14}\\
	&&q_1[J^{x-}]=\sum\limits_\lambda \left(
	j^{\lambda}(x^{\perp}, p_{\perp}, \partial_{x^{\perp}})
	+x \ h^{\lambda}(x^{\perp}, p_{\perp}, \partial_{x^{\perp}})\right) \Phi^{\lambda}(x^{\perp}),
	\label{2.15}\\
	&&q_1[J^{\bar x -}]=\sum\limits_\lambda \left(
	\bar j^{\lambda}(x^{\perp}, p_{\perp}, \partial_{x^{\perp}})
	+\bar x \ h^{\lambda}(x^{\perp}, p_{\perp}, \partial_{x^{\perp}})\right) \Phi^{\lambda}(x^{\perp}).
	\label{2.16}
\end{eqnarray}
We use the convention that $x^\perp$ arguments of  $h$, $j$ and $\bar j$ are  to the left from $\partial_{x^{\perp}}$.

\subsection{Kinematical Constraints}

Below we will solve for  $h$, $j$ and $\bar j$ by imposing (\ref{2.6}). We will first solve the kinematical constraints and then proceed to the dynamical ones. 

\subsubsection{Constraints on $h^{\lambda}(x^{\perp}, p_{\perp}, \partial_{x^{\perp}})$}
We start our analysis by solving the kinematical constraints for $q_1[P^-]$.
The consistency condition (\ref{2.6}) for $[P^-,P^x]=0$ leads to
\begin{equation}
	\label{4.1}
	[Q_2[P^-],q_1[P^x]]_\Phi+[q_1[P^-],Q_2[P^x]]_\Phi+
	[q_0[P^-],q_1[P^x]]_p +[q_1[P^-],q_0[P^x]]_p
	=0.
\end{equation}
Considering that $P^x$ is kinematical,  the first and the third terms in (\ref{4.1}) vanish. Simplifying the remaining terms, we arrive at
\begin{equation}
	\label{4.3}
	\sum\limits_{\lambda}\left(h^\lambda \cdot  \partial_{\bar x} \Phi^\lambda-\frac{\partial}{\partial \bar x}(h\Phi^{\lambda})\right)=0.
\end{equation}
It can be rewritten as 
\begin{equation}
	\label{4.4}
	\partial_{\bar x}h^{\lambda}=0,
\end{equation}
which, in turn, implies that the Hamiltonian kernel $h$ cannot depend on $\bar x$. This result is, of course, an anticipated consequence of translation invariance.

In a similar manner we consider commutators with other translations
\begin{equation}
	\label{4.5}
	[P^-,P^{\bar x}]=0 \qquad \Rightarrow \qquad \partial_x h^{\lambda}=0
\end{equation}
and
\begin{equation}
	\label{4.6}
	[P^-,P^{ +}]=0 \qquad \Rightarrow \qquad \partial_- h^{\lambda}=0.
\end{equation}
Consequently, $h^\lambda$ is independent of all $x^\perp$.
   
Next, we investigate the commutators of the Hamiltonian with kinematical Lorentz generators.  In particular, we find
\begin{equation}
	\label{4.7}
	[J^{x+},P^-]=0 \qquad \Rightarrow \qquad \sum\limits_{\lambda}\left(h^\lambda x\partial^+\Phi^\lambda - x\frac{\partial}{\partial x^-}(h^\lambda\Phi^\lambda)+p_-\frac{\partial}{\partial p_x}h^\lambda\Phi^\lambda\right)=0.
\end{equation}
In the second term $h$ can be trivially commuted with $\partial_-$, see (\ref{4.6}). Then, we further commute $h$ with $x$ in the second term and after cancellation with the first term, we find
\begin{equation}
	\label{4.8}
	\sum\limits_{\lambda}\left(\frac{\partial h^\lambda}{\partial(\partial_x)}\partial^+\Phi^\lambda +p_-\frac{\partial}{\partial p_x}h^\lambda\Phi^\lambda\right)=0.
\end{equation}
This yields a differential equation for $h^\lambda$  
\begin{equation}
	\label{4.9}
	\frac{\partial h^\lambda}{\partial(\partial_x)}\partial^+ +p_-\frac{\partial}{\partial p_x}h^{\lambda}=0.
\end{equation}

Analogously, one obtains the remaining kinematical constraints 
\begin{equation}
	\begin{split}
	\label{4.10}
	&[J^{\bar x+},P^-]=0 \qquad \Rightarrow \qquad 
	\frac{\partial h^{\lambda}}{\partial(\partial_{\bar x})}\partial^+ +p_-\frac{\partial}{\partial p_{\bar x}}h^{\lambda}=0, \\
	&[P^-,J^{+-}]=P^- \qquad \Rightarrow\qquad  \frac{\partial h^{\lambda}}{\partial (\partial^+)}\partial^++p^+ \frac{\partial}{\partial p^+}h^{\lambda}=-h^{\lambda},\\
	&[P^-,J^{x\bar x}]=P^- \qquad \Rightarrow 
	\qquad
	\partial_x \frac{\partial h^{\lambda}}{\partial (\partial_x)} 
	- \partial_{\bar x}\frac{\partial h^{\lambda}}{\partial (\partial_{\bar x})}+
	p_x \frac{\partial h^{\lambda}}{\partial p_x} 
	- p_{\bar x}\frac{\partial h^\lambda}{\partial p_{\bar x}}-\lambda h^{\lambda}=0.
	\end{split}
\end{equation}

We now proceed to solving (\ref{4.9}), (\ref{4.10}). Equation (\ref{4.9}) implies that $p_x$ and $\partial_x$ can enter $h^\lambda$ only in combination 
\begin{equation}
	\sigma_{ x}\equiv p_x-\partial_x \frac{p^+}{\partial^+}.
	\label{4.018}
\end{equation}
Analogously, the first equation from (\ref{4.10}) implies that  $p_{\bar x}$ and $\partial_{\bar x}$ can only enter $h^{\lambda}$ via 
\begin{equation}
	 \sigma_{ \bar x} \equiv p_{\bar x}-\partial_{\bar x} \frac{p^+}{\partial^+}.
	\label{4.018x1}
\end{equation}
Next,  the second equation in (\ref{4.10}) fixes the total homogeneity degree of $h^{\lambda}$ in $\partial^+$ and $p^+$ to be -1.
 Finally, the last equation in (\ref{4.10}) indicates that the total homogeneity degree of $h^{\lambda}$  in $\sigma_x$ and $\sigma_{\bar x}$ is $\lambda$.
Combining these constraints,  we find the general solution
\begin{equation}
	\label{4.12}
	h^\lambda = \frac{1}{p_-}\sigma_x^\lambda A^\lambda\left(\frac{\partial^+}{p_-},\sigma_x\sigma_{\bar x}\right),
\end{equation} 
where $A^\lambda$ is an arbitrary function of its arguments.

\subsubsection{Constraints on  $j^{\lambda}(x^{\perp}, p_{\perp}, \partial_{x^{\perp}})$ and $\bar j^{\lambda}(x^{\perp}, p_{\perp}, \partial_{x^{\perp}})$ }
We start by examining the kinematical constraint associated with
\begin{equation}
	\label{4.13}
	[J^{x-},P^{\bar x}]=-P^-.
\end{equation}
Due to our choice of convention for $q_1[J^{x-}]$ in (\ref{2.15}) -- more specifically, the shift by $xh$ -- $h$ cancels out on the both sides of equations like (\ref{4.13}). As a result, we find a simple equation that involves $j$ only 
\begin{equation}
	\label{4.14}
	\partial_{x}j^{\lambda}=0.
\end{equation} 
 Considering commutation relations with other translations, we obtain 
\begin{eqnarray}
	\label{4.15}
	[J^{x-},P^+]=P^x \qquad &\Rightarrow& \qquad\partial_{x^-}j^{\lambda}=0,\\
		\label{4.16}
	[J^{x-},P^x]=0 \qquad &\Rightarrow& \qquad\partial_{\bar x}j^{\lambda}=0.
\end{eqnarray}

Proceeding to commutators with the kinematical Lorentz generators, we find
\begin{eqnarray}
	\label{4.17}
	&&[J^{x-},J^{x+}]=0 \qquad \Rightarrow \qquad 
	\frac{\partial j^{\lambda}}{\partial(\partial_x)}\partial^+ +p_-\frac{\partial}{\partial p_x}j^{\lambda}=0,\\
&&	\label{4.18}
	[J^{x-},J^{\bar x +}]=-J^{-+}-J^{x\bar x}  \qquad \Rightarrow \qquad 
	\frac{\partial j^{\lambda}}{\partial(\partial_{\bar x})}\partial^+ +p_-\frac{\partial}{\partial p_{\bar x}}j^{\lambda}=0, \label{4.19}\\
&&	\label{4.20}
	[J^{x-},J^{+-}]=J^{x-}\qquad \Rightarrow\qquad  \frac{\partial j^{\lambda}}{\partial (\partial^+)}\partial^++p^+ \frac{\partial}{\partial p^+}j^{\lambda}=-j^{\lambda},\\
&&	\notag
	[J^{x-},J^{x\bar x}]=-J^{x-} \qquad \Rightarrow 
	\\
&& \qquad  \qquad\qquad	
\label{4.21}
	\partial_x \frac{\partial j^{\lambda}}{\partial (\partial_x)} 
	- \partial_{\bar x}\frac{\partial j^{\lambda}}{\partial (\partial_{\bar x})}+
	p_x \frac{\partial j^{\lambda}}{\partial p_x} 
	- p_{\bar x}\frac{\partial j^{\lambda}}{\partial p_{\bar x}}-(\lambda-1) j^{\lambda}=0.
\end{eqnarray}

The kinematical constraints (\ref{4.14})-(\ref{4.21}) are analogous to those for $h^\lambda$. Their general solution is 
\begin{equation}
	\label{4.22}
	j^\lambda = \frac{1}{p_-}\sigma_x^{\lambda-1} a^\lambda\left(\frac{\partial^+}{p_-},\sigma_x\sigma_{\bar x}\right),
\end{equation} 
where $a^\lambda$ is an arbitrary function of its arguments.

Similarly, we find 
\begin{equation}
	\label{4.23}
	\bar j^\lambda = \frac{1}{p_-}\sigma_{\bar x}^{-\lambda-1} \bar a^\lambda \left(\frac{\partial^+}{p_-},\sigma_x\sigma_{\bar x}\right),
\end{equation} 
where $\bar a^\lambda$ is yet another arbitrary function of its arguments.

\subsection{Dynamical constraints}

There are three dynamical constraints. The first one is associated with the commutation relation
\begin{equation}
	\label{4.24}
	[P^-,J^{x-}]=0.
\end{equation}
The second constraint is the complex conjugate of (\ref{4.24}) and, therefore, does not need to be studied separately. The last dynamical constraint ensures correct commutation of charges for $J^{x-}$ and $J^{\bar x-}$. It can be shown that it is always satisfied 
 as a consequence of the first two dynamical constraints, see e.g. \cite{Ponomarev:2016lrm}. It is, therefore, sufficient to derive the constraint associated with (\ref{4.24}).

Since, in (\ref{4.24}) both generators in the commutator are dynamical, all terms in (\ref{2.6}) are non-vanishing. Explicitly, we find
\begin{equation}
	\label{4.25}
	[Q_2[P^-],q_1[J^{x-}]]_\Phi+[q_1[P^-],Q_2[J^{x-}]]_\Phi+
	[q_0[P^-],q_1[J^{x-}]]_p +[q_1[P^-],q_0[J^{x-}]]_p
	=0.
\end{equation}
After some tedious, but straightforward computations, this leads to
\begin{eqnarray}
	&&j^{\lambda}
	\left(\frac{\partial_x{\partial_{\bar x}}}{\partial^{+}_x}+\frac{H}{p_-}\partial^{+}_x-\frac{\partial_{\bar x}p_{ x }+\partial_xp_{\bar x }}{p^-}\right)\nn
	&&-\frac{\partial h^{\lambda}}{\partial \partial_x}\frac{\partial_x{\partial_{\bar x}}}{\partial^{+}_x}-\frac{\partial h^{\lambda}}{\partial \partial^+_x}\partial_{\bar x}-H\frac{\partial}{\partial p_{ x}}
	h^{\lambda}-p_{\bar x} \frac{\partial}{\partial p_-}h^{\lambda}+h^{\lambda}\left(\lambda\frac{\partial_{\bar x}}{\partial^+_x}-\frac{p_{\bar x}}{p_-}\right)=0.
	\label{1.4}
\end{eqnarray}
Plugging here  $h$ and $j$ from (\ref{4.12}) and (\ref{4.22}), we obtain
\begin{eqnarray}
	s\left(\frac{m^2}{2}+u\right)a^{\lambda}(s, u)
	=-\left(u\ s\partial_s
	-\left(\frac{m^2}{2}+u\right)\left(\lambda+u\partial_u \right)\right)A^{\lambda}(s, u),
	\label{din}
\end{eqnarray}
where
\begin{equation}
\label{7jun1}
s\equiv \frac{\partial^+}{p_-}, \qquad u\equiv \sigma_{x}\sigma_{\bar x}.
\end{equation}
Quite remarkably, (\ref{din}) does not involve any other variables than $s$ and $u$.

\subsubsection{Locality}

In our analysis we have not yet fully accounted for constraints associated with locality. 
The standard way to impose locality in the light-cone approach is to require that the charges have certain analytic properties in transverse derivatives of fields. The precise choice of the assumptions one makes can be different, e. g. one may require that transverse derivatives enter charges only polynomially\footnote{Here ''transverse'' refers to $\partial^x$ and $\partial^{\bar x}$ only. Non-polynomial dependence in $\partial^+$ in the light-cone formalism does not imply that the theory is non-local. In particular, the Yang-Mills theory has negative powers of $\partial^+$ in the action.}. 
In fact, the precise choice of locality conditions is not necessary. As we will see below, the only condition that actually has effect on the solutions of the light-cone deformation procedure is the absence of certain combinations of transverse derivatives in the denominators of charges. 

To be more precise, we will first consider the deformation of particle's Hamiltonian (\ref{4.12}). The transverse derivatives enter $h$ only through $\sigma_x$ and $\sigma_{\bar x}$.
Accordingly, locality requires that these do not appear in the denominator of $h$. For $\lambda$ non-negative this means that $A^\lambda$ should not contain any $u$ dependence in the denominator, see (\ref{7jun1}). For $\lambda$ negative it is more convenient to represent $h^\lambda$ as
\begin{equation}
	\label{13jun1}
	h^\lambda = \frac{1}{p_-}\sigma_{\bar x}^{-\lambda} u^\lambda A^\lambda\left(\frac{\partial^+}{p_-},u\right).
\end{equation} 
Then, locality requires that $u^\lambda A^\lambda$ does not involve $u$ in the denominator. In a similar way locality is imposed on $j^\lambda$ and $\bar j^\lambda$.

Having clarified the constraints imposed by locality, we return to the analysis of (\ref{din}). We will consider the case of non-negative $\lambda$ first.

A trivial way of solving (\ref{din}) is to pick any $A$ and then solve for $a$, by simply dividing the both sides of the equation by 
$s(m^2/2+u)$. This way of solving (\ref{din}) is, however, not  satisfactory as it typically leads to $a$ with $(m^2/2+u)$ in the denominator and, thereby,  non-local $j$.

The general local solution of (\ref{din}) can be found as follows. The absence of $(m^2/2+u)$ in the denominator of $a$ is equivalent to saying that the left-hand side of (\ref{din}) vanishes for $u=-m^2/2$. Evaluating both sides of (\ref{din}) at $u=-m^2/2$, we find
\begin{equation}
\label{7jun2}
s\partial_s A^\lambda(s,u)\Big|_{u=-\frac{m^2}{2}}=0.
\end{equation}
This entails
\begin{equation}
\label{7jun12}
A^{\lambda}(s, u)=\mathcal{A}^\lambda(u)+\left(\frac{m^2}{2}+u\right)\mathcal B^\lambda(s, u),
\end{equation}
where both $\mathcal{A}$ and $\mathcal B$ are free of singularities in $u$.
The associated $a$ can then be found from (\ref{din})
\begin{equation}
 a^{\lambda}(s, u)= \frac{1}{s} \Big[(\lambda+u\partial_u)\mathcal A^\lambda(u)-u(s \partial_s-1) \mathcal B^\lambda(s, u)+\left(\frac{m^2}{2}+u\right)\left(\lambda+u\partial_u\right)\mathcal B^\lambda(s, u)\Big].
\end{equation}

The case of $\lambda$ negative can be studied analogously. Similarly, one should impose constraints resulting from locality of $\bar j^\lambda$.

\subsection{Fake interactions}
There are few reasons why non-linear corrections to the action may produce trivial interactions. In the field theory case this issue is well-known, see e.g. \cite{Berends:1984rq} for earlier discussions and \cite{Ponomarev:2022vjb} for a recent review. Below, we will extend this discussion to the case of a point particle in the field theory background and apply it to the problem we are dealing with. 

Below we will discuss fake interactions at the level of action. Considering that the action is related to the Hamiltonian  $q[P^-]$ in a simple way (\ref{6jun9}), this discussion can be easily translated to the language of charges that we employed before. Besides that,  we will often use the field theory terminology and call non-linear corrections to the point-particle action ''vertices''.

The first obvious way for a term in the action to give a trivial interaction  is to be a total derivative.

The second type of trivial point-particle vertices consists of vertices that are proportional to the field theory equations of motion. Indeed, we are discussing a regime in which a point particle does not affect the field theory background. This means that the latter satisfies the same field equations as if the point particle was absent. 
Accordingly, point-particle vertices, that are proportional to the field theory equations of motion vanish for on-shell backgrounds and do not change particle's dynamics.

Lastly, fake point-particle vertices can be obtained by field-dependent field redefinitions of particle's phase space variables. The standard argument shows, that these vanish once point particle's free equations of motion are taken into account. 

With this clarified, we will now find a class of trivial point-particle vertices, which are linear in fields. To start, we consider field's time derivative along a general trajectory of a  point particle
\begin{equation}
\label{7jun10}
	\frac{d\Phi^\lambda}{dx^+}=\frac{\partial \Phi^\lambda}{\partial x^+}  + \frac{\partial \Phi^\lambda}{\partial x}\dot x  + \frac{\partial \Phi^\lambda}{\partial \bar x}\dot{\bar x}+\frac{\partial \Phi^\lambda}{\partial x^-}\dot x^-.
\end{equation}
We then eliminate $\partial_+\Phi$ using the free equation of motion
\begin{equation}
\partial_+\partial_- \Phi^\lambda \approx - \partial_x \partial_{\bar x}\Phi^\lambda,
\end{equation}
where the symbol ''$\approx$'' indicates that this expression is only valid once the free equations of motion are taken into account.
In the full theory this equation has non-linear terms, but these do not contribute at the given order of perturbation theory. Next, we eliminate velocities in (\ref{7jun10}) in terms of momenta employing the free equations of motion of the point particle (\ref{7jun11}). We then find that
\begin{equation}
	\frac{d\Phi^\lambda}{d x^+}\approx -\frac{\partial^+}{p_-^2} \left(\sigma_x \bar\sigma_x +\frac{m^2}{2}\right)\Phi^\lambda.
	\label{4.36}
	\end{equation}

Let us now consider a point-particle vertex of the form
\begin{equation}
	S_{f}=	\int d x^+\ \alpha(x^\perp,p_\perp, \partial_{x^\perp})\left(\sigma_x \bar\sigma_x +\frac{m^2}{2}\right)\Phi^\lambda.
	\end{equation}
According to the above discussion, up to fake vertices it is equal to
\begin{equation}
	S_{f}\approx	-  \int d x^+ \alpha(x^\perp,p_\perp, \partial_{x^\perp})\frac{p_-^2}{\partial^+}\frac{d}{d x^+}\Phi^\lambda =
	 \int dx^+\frac{d}{dx^+} \left(\alpha(x^\perp,p_\perp, \partial_{x^\perp}) \frac{p_-^2}{\partial^+}\right)\Phi^\lambda.
	\end{equation}
This means that whenever a factor of $(u+m^2/2)$ is present in the action at the given order of perturbation theory it can be removed by adding fake interactions.

\subsection{Solution at the leading order}

As the last step of our analysis, we will  factor fake interaction out from the solution for $h$ that we found previously.

This is done as follows.
In (\ref{7jun12}) the $\mathcal B$ term  features a factor of $(u+m^2/2)$ explicitly, so we can fix the fake interaction ambiguity so that this term is absent. Similarly, all $u$ dependence in $\mathcal A$ can be removed, so that $\mathcal A$ gets replaced by some constant $C$. Eventually, this means that for non-negative helicity $\lambda$, the Hamiltonian of the point particle at the leading order in fields is given by
\begin{equation}
\label{13jun2}
q_1[P^-]= C^{\lambda}\frac{\sigma^{\lambda}_x}{p_-} \Phi^{\lambda}(x^{\perp}), \qquad \lambda \ge 0,
\end{equation}
where $C^\lambda$ is an arbitrary coupling constant or, equivalently, a charge responsible for the interaction of a particle with the helicity-$\lambda$ field.

For negative helicities one proceeds analogously, except that to implement locality it is more convenient to use the representation (\ref{13jun1}).
Combining the results for all helicities, we find the final formula for the leading order correction to the Hamiltonian 
\begin{equation}
\label{9junx1}
	H_p=q_1[P^-]=\sum\limits_{\lambda} \left(C^{\lambda}\frac{\sigma^{\lambda}_x}{p_-}+\bar C^{-\lambda}\frac{\sigma^{-\lambda}_{\bar x}}{p_-}\right) \Phi^{\lambda}(x^{\perp}),
	\end{equation}
where $C^\lambda$ and $\bar C^{\lambda}$ are arbitrary coupling constants satisfying
\begin{equation}
\label{13jun4}
C^{\lambda}=0, \qquad \lambda<0 \qquad \text{and} \qquad \bar C^{-\lambda}=0, \qquad \lambda \ge 0. 
\end{equation}
The associated corrections to the dynamical Lorentz generators are
\begin{eqnarray}
		&&q_1[J^{x-}]=\sum\limits_\lambda \left(
		\lambda C^{\lambda}\frac{\sigma^{\lambda-1}_x}{\partial^{+}}
		+x \left(C^{\lambda}\frac{\sigma^{\lambda}_x}{p_-}+\bar C^{ -\lambda}\frac{\sigma^{-\lambda}_{\bar x}}{p_-}\right) \right) \Phi^{\lambda}(x^{\perp}),\\
		&&q_1[J^{\bar x -}]=\sum\limits_\lambda \left(-\lambda \bar  C^{-\lambda}\frac{\sigma^{-\lambda-1}_{\bar x}}{\partial^{+}}
		+\bar x \left(C^{\lambda}\frac{\sigma^{\lambda}_x}{p_-}+\bar C^{ -\lambda}\frac{\sigma^{-\lambda}_{\bar x}}{p_-}\right) \right) \Phi^{\lambda}(x^{\perp}).
	\end{eqnarray}

In appendix \ref{App2} we compared our result (\ref{9junx1}) with the result of gauge fixing of the covariant vertices at this order \cite{deWit:1979sib}
and with the coupling constants fixed as in \cite{Segal:2000ke}.
We found that gauge fixing of \cite{deWit:1979sib} leads to the same set of vertices (\ref{9junx1}). In addition, the coupling constants found in \cite{Segal:2000ke} translate into 
\begin{equation}
\label{13jun5}
C^\lambda=e, \qquad \bar C^{-\lambda}=e
\end{equation}
for all non-vanishing couplings  (\ref{13jun4}). In (\ref{13jun5}) $e$ is a single coupling constant. It would be interesting to verify whether the same dependence holds true for chiral theories.

\section{Conclusion}
\label{sec:5}

In the present paper we developed the light-cone formalism for a point particle in a higher-spin background. Following the familiar logic of the light-cone approach, we started from a free particle in  Minkowski space and derived the associated phase space generators of the Poincare algebra. These generators receive corrections once the particle is coupled to a non-trivial background. By requiring that the deformed generators still satisfy the commutation relations of the Poincare algebra, one finds a set of constraints that these generators have to satisfy. 
Considering that the Hamiltonian is just the generator of translations in time, the point particle action in the  Hamiltonian form can be readily given. In this analysis we encountered some peculiarities, which are not characteristic for the field theory case. In particular, we explained how back-reaction of a point particle to the background  -- which is known to lead to divergences of self-energy type -- can be consistently removed. 

We applied this procedure for deriving the correction to the point particle action at the first non-trivial order, that is when higher-spin fields appear linearly. At this order interactions of higher-spin fields are not yet relevant. We found that there is a single consistent coupling of a point particle to a higher-spin field of each helicity. Each such interaction has an independent coupling constant\footnote{This coupling constant does not have to be universal with universality understood as in Weinberg's soft graviton theorem \cite{Weinberg:1964ew}. Namely, another point particle may couple to the same higher-spin field with a different coupling constant. This conclusion may, in principle, change once the analysis is extended to higher orders in perturbations.}. This result is reminiscent of the analogous one in the field theory: there is a unique cubic vertex that involves a massless field of a given helicity linearly and a scalar field quadratically \cite{Bengtsson:1983pd,Bengtsson:1986kh}, moreover, the leading order analysis leaves the associated coupling constant unconstrained.

We also compared our results with the available ones in the covariant approach \cite{deWit:1979sib}. We found that once the light-cone gauge fixing of the covariant result \cite{deWit:1979sib} is carried out the latter matches the action that we found in the light-cone formalism. Thus, unlike in the field theory case, in which covariant and light-cone classifications of vertices disagree already at the cubic order \cite{Bengtsson:2014qza,Conde:2016izb}, we found that, at least, at the given order the two approaches give equivalent results. 

It would be interesting to extend our results to higher orders in higher-spin fields. At higher orders interactions of higher-spin fields become relevant. 
The most natural interacting higher-spin theory to use in this context is the chiral higher-spin theory  \cite{Metsaev:1991mt,Metsaev:1991nb,Ponomarev:2016lrm}. We expect that already at the next order in fields, the action of the point particle becomes severely constrained. More precisely, it may happen that it does not exist at all. Alternatively, if it does exist, it seems natural to expect that the higher-spin couplings of a point particle will be fixed almost completely in terms of the coupling constants of the higher-spin theory. It would be also interesting to see whether these higher-spin charges agree with those, we extracted by gauge fixing \cite{Segal:2000ke}, the latter result relying on the conformal higher-spin symmetry. Ultimately, it would be interesting to obtain the point particle action to all orders in higher-spin fields 
as a closed-form expression
 and use it to study geometric features of chiral higher-spin backgrounds.

\acknowledgments

We would like to thank  E. Skvortsov  for comments on the draft.

	\appendix
	\section{Notations \label{App}}

We use the mostly plus convention for  the 4d Minkowski metric
\begin{equation}
		ds^2 = -(dx^0)^2+ (dx^1)^2+(dx^2)^2+(dx^3)^2.
	\end{equation}
	The light-cone coordinates are defined by
	\begin{align}
		\notag
		x^+ &\,= \frac{1}{\sqrt{2}}(x^3+x^0), & x^-&\, = \frac{1}{\sqrt{2}}(x^3-x^0),\\
		\label{29sep1}
		x &\,=\frac{1}{\sqrt{2}}(x^1-ix^2), & \bar x &\,= \frac{1}{\sqrt{2}}(x^1+ix^2).
	\end{align}
	In these coordinates the metric becomes off-diagonal 
	\begin{equation}
		ds^2 = 2dx^+ dx^- + 2 dx d\bar x.
	\end{equation}
For derivatives the coordinate transformation (\ref{29sep1}) leads to
\begin{align}
		\notag
		\partial^-  &\,= \frac{1}{\sqrt{2}}(\partial^3-\partial^0), &  \partial^+ &\, = \frac{1}{\sqrt{2}}(\partial^3 + \partial^0),\\
		\bar\partial &\,= \frac{1}{\sqrt{2}}(\partial^1-i\partial^2), &  \partial &\, =\frac{1}{\sqrt{2}}(\partial^1 + i \partial^2).
	\end{align}
In particular,
	\begin{equation}
		\partial^+ x^- = \partial^- x^+ = \bar\partial x = \partial\bar x = 1.
	\end{equation}

	\section{Comparison with covariant results \label{App2}}

In this appendix we will compare the covariant results of \cite{deWit:1979sib,Segal:2000ke} with the results obtained in our paper. More specifically, we will use the leading order action from \cite{Segal:2000ke}, which in comparison to \cite{deWit:1979sib} also has the relative couplings of a point particle to higher-spin fields fixed. 

The action we start from is 
\begin{equation}
\label{8jun1}
S = -m \int d\tau \sqrt{-\eta_{\mu\nu}\dot x^{\mu} \dot x^{\nu}}\left(1+\frac{e}{m^2}\sum_{s=0}^\infty \varphi_{m_1\dots m_s}\dot x^{m_1}\dots \dot x^{m_s}\left(\frac{m}{\sqrt{-\eta_{\mu\nu}\dot x^{\mu} \dot x^{\nu}}} \right)^s  \right),
\end{equation}
where $e$ is a single coupling constant, while $\varphi$ is the totally symmetric Fronsdal field \cite{Fronsdal:1978rb}. Our first task is to impose the light-cone gauge on higher-spin fields. 

In the light-cone gauge one has
\begin{equation}
\label{8jun2}
\varphi^{+m_2\dots m_s}=0.
\end{equation}
The non-vanishing components of the Fronsdal fields are given by, see e. g. \cite{Ponomarev:2022vjb},
\begin{equation}
\label{8jun3}
\begin{split}
\varphi^{-(n)x(s-n)}&=\left(-\frac{\partial_x}{\partial^+} \right)^n \Phi^s, \qquad 
\varphi^{-(n)\bar x(s-n)}=\left(-\frac{\partial_{\bar x}}{\partial^+} \right)^n \Phi^{-s}, \\
\varphi^{-(s)}&= \left(-\frac{\partial_x}{\partial^+} \right)^s \Phi^s +\left(-\frac{\partial_{\bar x}}{\partial^+} \right)^s \Phi^{-s}.
\end{split}
\end{equation}
Here, e. g., notation ''$-(n)x(s-n)$'' means that $n$ indices of a tensor take value ''$-$'' and the remaining $s-n$ indices take value ''$x$''. The Fronsdal field is symmetric, so (\ref{8jun3}) is valid irrespectively of the order of indices of tensors on the left-had side. Overall, there are $\frac{s!}{n!(n-s)!}$ components of a tensor $\varphi$ with $n$ indices taking value ''$-$'' and the remaining $s-n$ indices take value ''$x$''.

Relations (\ref{8jun3}) hold for the free theory. In the non-linear theory these receive corrections, which are, however, irrelevant at the order of perturbation theory we are dealing with. 
It also needs to be remarked that conformal higher-spin theories have the spectrum, which besides massless degrees of freedom contains other non-unitary excitations. In the present context we are interested only in the massless sector of conformal higher-spin theories and the associated couplings of the point particle.

Utilizing (\ref{8jun3}), we  find
\begin{equation}
\label{8jun4}
\begin{split}
\varphi_{m_1\dots m_s}\dot x^{m_1}\dots \dot x^{m_s} =\sum_{n=0}^s \frac{s!}{n!(n-s)!}\left[\left(-\frac{\partial_x}{\partial^+} \right)^n \Phi^s (\dot{\bar{x}})^{s-n}+
\left(-\frac{\partial_{\bar x}}{\partial^+} \right)^n \Phi^{-s} (\dot{{x}})^{s-n}
\right]\\
=\left(\dot{\bar{x}} - \frac{\partial_x}{\partial^+} \right)^s\Phi^s+\left(\dot{{x}} - \frac{\partial_{\bar x}}{\partial^+} \right)^s\Phi^{-s}.
\end{split}
\end{equation}
Plugging this into (\ref{8jun1}) we obtain
\begin{equation}
\label{8jun5}
\begin{split}
S=&-m\int dx^+ \sqrt{-2\dot x^- -2\dot x \dot{\bar x}} \Big[ 1+\frac{e}{m^2}\Phi^0\\
&+\frac{e}{m^2}\sum_{\lambda =1} 
\left(\frac{m}{\sqrt{-2\dot x^- -2\dot x \dot{\bar x}}} \right)^\lambda \left(
\left(\dot{\bar{x}} - \frac{\partial_x}{\partial^+} \right)^\lambda\Phi^\lambda+\left(\dot{{x}} - \frac{\partial_{\bar x}}{\partial^+} \right)^\lambda\Phi^{-\lambda}
 \right)\Big].
\end{split}
\end{equation}

To compare (\ref{8jun5}) with our result, we need to perform the Legendre transform. In general, one has
\begin{equation}
\label{9jun1}
p_i \equiv \frac{\partial L}{\partial \dot q^i}(q,\dot q, \Phi).
\end{equation}
Let the solution of (\ref{9jun1}) for $\dot q$ in terms of $p$ be 
\begin{equation}
\label{9jun2}
\dot q^i = \dot q^i(p,q,\Phi).
\end{equation}
Then, the Hamiltonian is
\begin{equation}
\label{9jun3}
H(p,q,\Phi)\equiv p_i \dot q^i(p,q,\Phi)-L(q,\dot q(p,q,\Phi),\Phi).
\end{equation}

At a given order of perturbation theory, we are interested in evaluating $H$ up to linear terms in fields. Decomposing both sides of (\ref{9jun3}) to this order in $\Phi$, we obtain
\begin{equation}
\label{9jun4}
\begin{split}
H(p,q,0)&+ \frac{\partial H}{\partial \Phi}(p,q,\Phi)\Big|_{\Phi=0}\Phi+O(\Phi^2) \\
&= p_i \dot q^i(p,q,0)-L(q,\dot q(p,q,0),0)
\\&+
p_i\frac{\partial \dot q^i}{\partial \Phi}(p,q,\Phi)\Big|_{\Phi=0}\Phi
-\frac{\partial L}{\partial \Phi}(q,\dot q(p,q,\Phi),\Phi)\Big|_{\Phi=0}\Phi
\\&-
\frac{\partial L}{\partial \dot q^i}(q,\dot q(p,q,\Phi),\Phi)\Big|_{\Phi=0} \frac{\partial \dot q^i}{\partial \Phi}(p,q,\Phi)\Big|_{\Phi=0}\Phi+O(\Phi^2).
\end{split}
\end{equation}
It follows from (\ref{9jun1}) that
\begin{equation}
\label{9jun5}
p_i = \frac{\partial L}{\partial \dot q^i}(q,\dot q(p,q,\Phi),\Phi)\Big|_{\Phi=0} +O(\Phi).
\end{equation}
Substituting this into (\ref{9jun4}), at the linear order in $\Phi$ we find
\begin{equation}
\label{9jun6}
\begin{split}
 \frac{\partial H}{\partial \Phi}(p,q,\Phi)\Big|_{\Phi=0}\Phi
= 
-\frac{\partial L}{\partial \Phi}(q,\dot q(p,q,\Phi),\Phi)\Big|_{\Phi=0}\Phi.
\end{split}
\end{equation}
In other words, at a given order, the Hamiltonian is minus the Lagrangian for which the velocities should be eliminated in terms of momenta using the free theory expressions. Indeed, as we showed, the derivatives of $\dot q$ in $\Phi$ cancelled out from the right-hand side of (\ref{9jun4}), thus, the relation (\ref{9jun2}) at the given order can be replaced with the free theory one.

Accordingly, from (\ref{9jun6}), (\ref{8jun5}) and (\ref{7jun11}) we find
\begin{equation}
\label{9jun7}
H_3 = \frac{e}{p_-}\Phi^0 + \frac{e}{p_-}\sum_{\lambda = 1}^{\infty}(\sigma_x^\lambda\Phi^\lambda+\sigma^\lambda_{\bar x}\Phi^{-\lambda}).
\end{equation}

\bibliography{pp}
\bibliographystyle{JHEP}

\end{document}